\documentclass[aps,prl,twocolumn,nofootinbib,epsfig]{revtex4}
\usepackage{amsmath,amssymb,latexsym}
\usepackage{graphicx}
\usepackage{bm}
\def\be{\begin{equation}}
\def\ee{\end{equation}}
\def\ed{\end{document}}

\begin{document}

\title{ On the determination of $\cal CP$-even and $\cal CP$-odd components of a mixed $\cal CP$ Higgs boson at $e^{+}e^{-}$ linear colliders  }
\author{Mar\'{\i}a Teresa Dova and Sergio Ferrari}
\affiliation{Departamento de F\'{\i}sica, Universidad Nacional de La Plata,
C.C.67 -- 1900 La Plata, Argentina}

\begin{abstract}

We present a method to investigate the $\cal CP$ quantum numbers of the Higgs boson in the process $e^+ e^- \rightarrow Z \phi$ at a future $e^{+}e^{-}$ linear collider (LC), where $\phi $, a generic Higgs boson, is a mixture of $\cal CP$-even and $\cal CP$-odd states. The procedure consists of a comparison of the data with predictions
obtained from Monte Carlo simulations corresponding to the productions of scalar and pseudoescalar Higgs and the interference term which constitutes a distinctive signal of $\cal CP$ violation.  We present estimates of the sensitivity of the method from Monte Carlo studies using hypothetical data samples with a full LC detector simulation taking into account the background signals.

\end{abstract}
\pacs{14.80.Bn;14.80.Cp -- Keywords: Higgs bosons, Linear Colliders}

\maketitle

\section{Introduction}
The future linear $e^+$ $e^-$ collider TESLA  is planned to work with
a maximum center-of-mass energy of 500 GeV, extendable to 800 GeV without modifying the original design
\cite{DR}. It will have a luminosity of $3.4\times 10^{34}$ $cm^{-2}$ $s^{-1}$, a thousand times greater than the LEP at CERN, and so it will
be well suited for a discovery of a light Higgs boson. Even if
the Higgs is discovered before at Tevatron (Fermilab) \footnote{ The new world average of the expected Higgs mass of 117 GeV \cite{d0} is yet accesible in the current run of the Tevatron} or at the future LHC (CERN, Geneva), $e+e-$ colliders are the ideal machines to investigate
the Higgs sector in the intermediate mass range since all
major decay modes can be explored, with the Higgs particle
produced through several mechanisms \cite{A1}. For a light
or intermediate mass Higgs boson, the Higgstrahlung
process $e^+ e^- \rightarrow Z\Phi$, where $\Phi$ denotes a generic Higgs
boson, is expected to be
the most promising process to study its properties and
interactions and to search for deviations from the Standard
Model (SM) predictions(see \cite{P2} and references therein). A comprehensive review
of the Higgs boson properties has been given in ref.
\cite{kniehl}. The theory of the Higgs bosons, with emphasis on
the Higgs scalars of the SM and its non-supersymmetric
and supersymmetric extensions has been recently presented
in ref.\cite{gunion}.  The spin, parity and charge conjugation
quantum numbers, $J^{PC}$, of the Higgs boson can
potentially be determined independently of the model.
It has been shown  that measurements  of the
threshold dependence of the Higgsstrahlung cross-section
constrains the possibles values $J^{PC}$ of the state \cite{TD}. In
the minimal Standard Model the Higgs mechanism requires only one Higgs doublet to generate masses for fermions and gauge bosons \cite{H}. It leads to the appearence of a neutral $\cal CP$-even Higgs ($H$). In the two-doublet Higgs model (2DHM) or the supersymmetric extension of the SM \cite{HH}, neglecting $\cal CP$ violation, there are two $\cal CP$-even
states ($h, H$) and one $\cal CP$-odd state ($A$), plus a pair of
charged Higgs bosons ($H^{\pm}$).  In a general 2DHM the three neutral Higgs bosons could correspond to arbitrary mixtures of $\cal CP$ states and their production and decay exhibits $\cal CP$ violation. The angular distributions of
the Higgsstrahlung cross section depends upon whether
the $\Phi$ is $\cal CP$-even, $\cal CP$-odd, or a mixture \cite{P2, DK, P1,cp2, cp3}. Also
the angular distribution of the fermions in the $Z \rightarrow f \bar f$
from $Z\Phi$ production reflects the $\cal CP$ nature of the state $\Phi$
\cite{cp2,P2, DK,kniehl1}.  An analysis of the angular distributions
of the final state fermions in the Higgsstrahlung process
with the formalism of optimal variables has been performed
in ref.\cite{MS2}. A fit to double-differential angular distribution in the production and decay angles results in a clean separation between a scalar and pseudoscalar states assuming that 
the $Z\Phi$ cross section is independent of the $\cal CP$ nature of the $\phi$ \cite{derwent}. Recently, the prospects for the meaurement of the pseudoscalar admixture in the $h\tau \tau$ coupling to a SM Higgs boson was presented \cite{was}.  

In this paper, we present an alternative  method that simultaneouly
uses the distributions of the production and decay
angles to distinguish the SM-like Higgs boson from a $\cal CP$-odd
$0^{-+}$ state $A$, or a $\cal CP$-violating mixture $\Phi$. We perform an analyisis of Monte Carlo events that takes into account the signals and background, as well as a simulation of a TESLA detector response. In the next section we shall present the theoretical
ansatz considered and the  details of the Monte Carlo
simulation used to generate the $e^+ e^- \rightarrow Z\Phi$ samples.
We then describe the proposed method, detector simulations and the imposed cuts for the event
selection. Finally we present the fit techniques and the results obtained from Monte Carlo studies.       

\section{$e^+ e^- \rightarrow Z \Phi$ samples}

Events of the signal $e^+ e^- \rightarrow Z H $ were generated using the
{\sc Pythia} program \cite{PY}. The cross section for the Higgsstrahlung process is given by:
\begin{equation}
\sigma(e^+ e^- \rightarrow ZH) = \frac{G_F^2 M_Z^4}{96\pi s}(a_e^2+v_e^2)\beta
\frac{\beta^2 + 12 M_Z^2/s}{(1- M_Z^2/s)^2}
\end{equation}

with $\beta = \sqrt{[s - (M_H + M_Z)^2][s - (M_H - M_Z)^2]}/s$.

The effects of initial state bremstrahlung were included in the {\sc Pythia} generation. 

For the production of a Higgs boson $\Phi$ with arbitrary
$\cal CP$ properties, $e^+ e^- \rightarrow Z \Phi$,  the amplitude for the process
(which is not included in {\sc Pythia}) can be described
by adding a ZZA coupling with strength $\eta $ to the SM
matrix element as in ref.\cite{DR}. The matrix element of
the process containing both the $\cal CP$ even amplitude, ${\cal M}_{ZH}$,
 and a $\cal CP$ odd amplitude, ${\cal M}_{ZA}$,  is given by:

\begin{equation}
{\cal M}_{Z\Phi} = {\cal M}_{ZH} + \imath \eta \cdot {\cal M}_{ZA}
\end{equation}
where $\eta$ is a dimensionless factor. The total cross section
depends of the value of $\eta$ as follows \cite{MS2}:

\begin{equation}
\sigma(\eta,s) = \frac{G_F^2 M_Z^6\beta}{16\pi}\frac{1}{D_Z(s)}(v_e^2+a_e^2)(2 +
\frac{s\beta^2}{6M_Z^2} +\eta^2 \frac{s^2\beta^2}{M_Z^4})
\label{eta1}
\end{equation}

where
\begin{eqnarray}
D_{Z}(s) & = & (s-M_{Z}^{2})^2 + M_{Z}^{2}\Gamma_{Z}^2
\end{eqnarray}

and $M_{Z}$, $\Gamma_{Z}$ denote the $Z$ boson mass and width, $G_F$
is the Fermi constant, and $v_e, a_e$ are the usual vector and
axial-vector coupling constants of the $e$ to the $Z$ boson. In the SM $\eta$ is
zero. In the MSSM a ZZA is forbidden at Born level,
but is induced via higher-order loop effects \cite{HH}. In general in
extensions of the Higgs sector, $\eta$ need  not to be loop suppressed, and may be arbitrarily large. Hence, it is useful to allow for $\eta$ to be a free parameter in the data analysis.

\begin{figure}
\begin{center}
\includegraphics[height=4.cm]{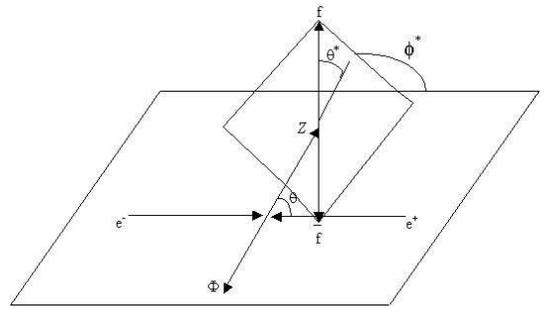} 
\caption{Definition  of the production and decay angles of the process $e^+ e^- \rightarrow Z\Phi [Z \rightarrow f \overline{f}$].}\label{fig1} 
\end{center}
\end{figure}

As it was mentioned in the Introduction, the quantum numbers
$J^{PC}$ of the Higgs bosons can be determined
at future $e^+ e^- $ linear colliders in a model independent
way by analysing the angular dependence of the Higgstrahlung
process. The most sensitive kinematic variable to distinguish the different contributions to Higgs boson production is  $\theta$, the polar angle
of the $Z$ boson w.r.t. the beam axis in the laboratory
frame.  The sensitivity can be increased by including the
angular distributions of the decay to fermions, $Z\rightarrow f \bar{f}$ in the boson rest
frame (See Fig.~\ref{fig1}). Here the z-axis is chosen along the
direction of the Z-boson momemtum.
The decay amplitude is then a function of the angle between the Z momentum and $f$, $\theta^*$, and the angle between the Z production plane and the Z decay plane, $\phi^*$. 

To obtain the angular distributions corresponding
to the non-SM Higgs $\Phi$ with arbitrary $\cal CP$ properties in the process
$e^+ e^- \rightarrow Z\Phi$, we have used a ``re-weighting'' method.
This procedure allows one to obtain the distributions for  arbitrary values of $\eta$ by weighting the distributions for $\eta = 0$ according to the differential cross section in $\theta, \theta^*, \phi^*$. The
weight factor is given by the following ratio:

\begin{equation}
W(\cos \theta, \cos \theta^*, \cos \phi^*)= \frac{|{\cal M}_{Z\Phi}(\eta)|^2}{|{\cal M}_{ZH}|^2}
\end{equation}

\begin{figure}
\centering 
\includegraphics[height=8.cm]{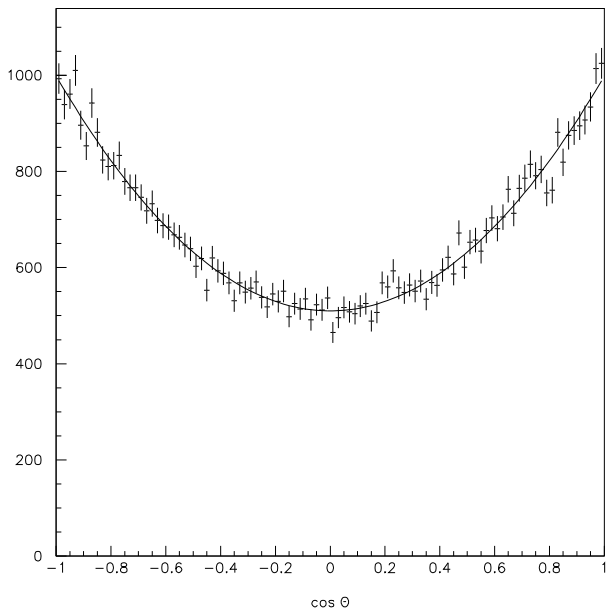}
\caption{Angular distribution of the process  $e^+ e^- \rightarrow Z A$  obtained by the ``re-weighting'' method (See text) for an integrated luminosity of  500 $fb^{-1}$ and a center of mass
energy of $\sqrt{s}=350$ GeV, assuming a Higgs mass of 120 GeV. The line indicates the exact theoretical dependence. }\label{za} 
\end{figure}

The squared amplitude $|{\cal  M}_{Z\Phi}(\eta)|^2$  has three contributions:
\begin{equation}
|{\cal M}_{Z\Phi}(\eta)|^2 = |{\cal M}_{ZH}|^2 + \eta \cdot 2\Im m({\cal M}^*_{ZH}{\cal M}_{ZA}) + \eta^2 |{\cal M}_{ZA}|^2
\end{equation}
The first term reproduces the SM-like cross section. The interference term between the
$\cal CP$ even and $\cal CP$ odd amplitudes, linear in $\eta$, generates a forward-backward asymmetry, that is a hallmark of $\cal CP$ violation. The third term correspond to the pseudoescalar
Higgs cross section. Of course, $\eta=0$ brings us
back to the scalar SM Higgs production. The explicit expression for the
squared amplitude of the bremsstrahlung process in terms of  $\theta$, $\theta^*$, $\phi^*$ is taken from
ref. \cite{DK}.  In our procedure, the weight is re-scaled to be lower than 1,
for a better treatment of errors. To check the reliability
of the method we compared the obtained distributions
using Monte Carlo with the analytical expressions. 
Figure ~\ref{za} shows the obtained production angular distribution for the process $e^+ e^- \rightarrow ZA$ using the procedure 
described above along with the analytical form. 
The distribution is in very good
agreement with the theoretical expectation proving the
validity of the ``reweighting'' procedure.

\section{Description of the method and Monte Carlo studies}
 
We consider the production of Higgs events
at the TESLA operating at a center-of-mass energy of 350
GeV, assuming an integrated luminosity of 500 $fb^{-1}$. At this energy the main production
process for the Higgs boson in the SM is the
Higgsstrahlung process, $e^+ e^- \rightarrow ZH$ \cite{JE}. 
The corresponding expected number
of events for this process is $6.6\times 10^4$.

We have choosen for the present study the process
$e^+ e^- \rightarrow ZH \rightarrow  \mu^+ \mu^{-}H$ with a Higgs boson mass of
120 GeV. This decay channel exhibits a clean signature in
the detector and the selection efficiencies are
expected to be independent of the decay mode of the
Higgs boson.   We allow the produced Higgs to be  either  scalar
 or a mixture state $\Phi$ including an interference term.

All the Monte Carlo
samples have been generated with the {\sc Pythia} program as described in the previous section. These events are then passed through the simulation package SIMDET \cite {SD}, a parametric Monte Carlo program
for a TESLA detector \cite{SD2} which follows the proposal presented in the TESLA Conceptual Design Report \cite{DR}. For the Higgs boson all decay modes are simulated as expected in the SM. The following background processes are considered in the analysis: $e^{+}e^{-} \rightarrow e^{+}e^{-}f^{+}f^{-}$, $e^{+}e^{-} \rightarrow f^{+}f^{-}(\gamma)$,  $e^{+}e^{-} \rightarrow W^{+}W^{-}$ and  $e^{+}e^{-} \rightarrow ZZ$.  Both signal and background
events are processed by the detector simulation package.  

For the event selection we follow ref. \cite{MS2}.
At least one muon and anti-muon are identified, with energy larger than 15 GeV.
The mass of the di-muon system is required to be consistent with the Z boson
hypothesis within 5 GeV.
The recoil mass of the di-muon system $M^{2}_{rec} = (\sqrt{s} - (E_{\mu^+} + E_{\mu^-}))^{2} - (\vec P_{\mu^+} + \vec P_{\mu^-})^{2}$ has to be consistent with the H
boson hypothesis within 5 GeV. This variable will yield a peak for the signal of the Higgs boson mass, independently of the Higgs boson decay mode. 
To remove a significant part of the remaining background, the absolute z-component of the di-muon system is required to be 
smaller than 120 GeV.

\begin{figure}
\centering
\includegraphics[height=8.cm]{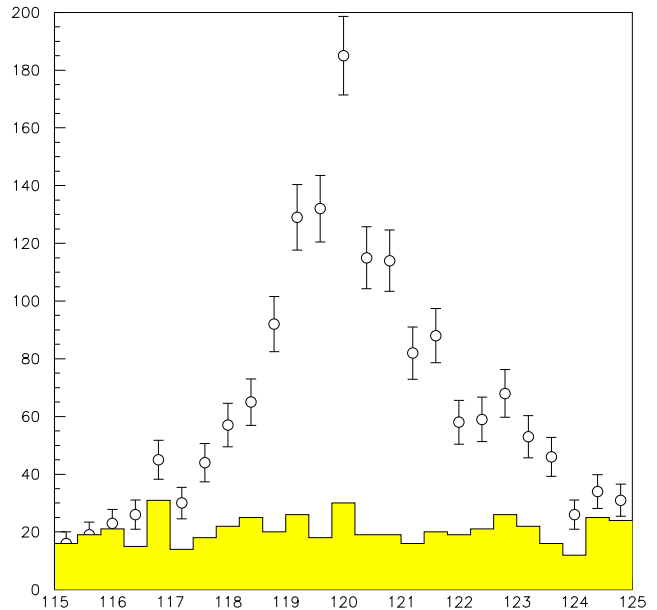}  
\caption{Recoil mass spectra off the Z in $e^+ e^- \rightarrow Z H \rightarrow \mu^+ \mu^- X$. Shadowed area represents the expected background events.}\label{Recoil} 
\end{figure}

\begin{figure}
\centering
\includegraphics[height=8cm]{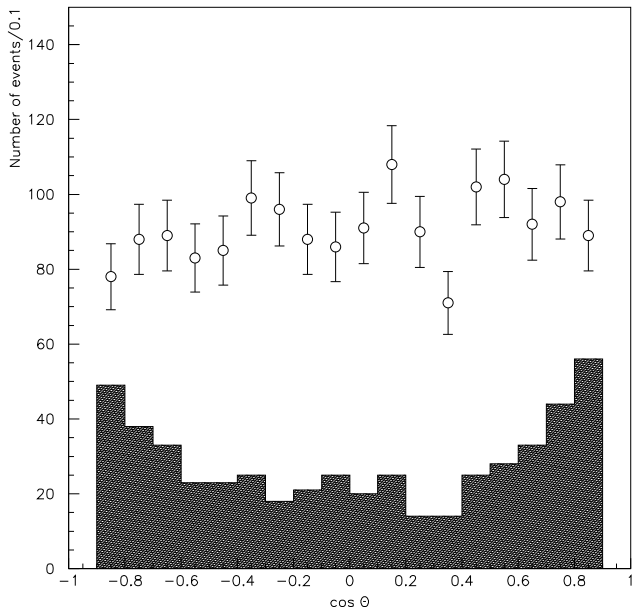} 
\caption{Angular distribution, cos $\theta$, of the selected events in $e^+ e^- \rightarrow Z H \rightarrow \mu^+ \mu^- X$. The shaded histogram correspond to the expected background.}\label{angular} 
\end{figure}

The momemtum of the selected muons are used to calculate the cosines of the production 
and decay angles for
futher use in the method to determine the $J^{PC}$ properties of the Higgs boson.
 It has been noted in \cite{DR} that having excellent momemtum and energy resolution will allow the Z to be well reconstructed. The recoil mass against the Z, can then be used to detect the Higgs boson and to study its properties.
 Figure~\ref{Recoil} shows the recoil mass distribution for the $m_{H} = 120$ GeV signal, obtained from the selected events in the sample of $e^+ e^- \rightarrow Z H \rightarrow \mu^{+} \mu^{-}X $. The Higgs boson signal appears on top of a small background. In Figure ~\ref{angular}  the corresponding $\cos \theta$ distribution is shown. The expected background is also presented. The combination of the cut on the z-component of the di-muon system, and the decreasing muon identification performance results in an efficiency for 
$\cos \theta > 0.9$ close to zero.

The kinematics of the   $e^+ e^- \rightarrow Z\Phi$ $[Z \rightarrow f \overline{f}]$ process is
described by the production and decay angles $\theta$, $\theta^*$, $\phi^*$. 
The method we propose consists in generating 3-dimensional distributions in $\cos \theta$, $\cos \theta^*$ and $\cos \phi^* $ using the Monte Carlo events generated as described above for each contribution in equation (3). We write then the likelihood: 

\begin{equation}
{\cal L} = \prod_{(\cos \theta)_i, (\cos \theta^*)_j, (\cos \phi^*)_k}  
\frac{\mu_{ijk}{}^{N_{data}(i,j,k)} e^{-\mu_{ijk}}}{N_{data(i,j,k)}!} 
\end{equation}

where $N_{data}(i,j,k)$ is the number of events of the hypothetical data 
sample and $\mu_{ijk}$ is the expected number in the ijk-th bin.  $\mu_{ijk}$ 
is calculated assuming a linear combination of the number of events of
three Monte Carlo samples, corresponding to the production of scalar Higgs
(MC\_ZH), pseudoscalar (MC\_ZA) Higgs and  events for the interference term
(MC\_IN):
\begin{equation}
\mu_{ijk} = {\cal N}.(\alpha.MC\_ZH_{ijk} + \beta.MC\_IN_{ijk} + \gamma.MC\_ZA_{ijk})
\end{equation}
where $\cal N$ is the overall normalization factor between numbers of data
and Monte Carlo events which can be fixed (${\cal N}=0.1$ in our case) 
or left free as a further check of the fit.
  The likelihood is then maximized with respect to $\alpha$, $\beta$ and
$\gamma$. The absolute value of $\beta$ indicates the contribution of interference term in the sample and $\alpha$ and $\gamma$ indicate the fraction of scalar and pseudoscalar components respectively. A significant deviation of  $\beta$ from zero would imply the existence of $\cal CP$ violation, independent of the specific model.  For a scalar Higgs sample ($\eta=0$), the result of the fit is expected to be $\alpha=1$ and $\beta=\gamma=0$. 

We have performed  Monte Carlo studies with several  hypothetical data samples with non-standard values of $\eta$.  A maximum likelihood fit for the best linear combination of MC\_ZH, MC\_ZA and MC\_IN to match the hypothetical data sample gave statistical errors of 0.04, 0.02 and 0.04 for $\alpha$, $\beta$ and $\gamma$, respectively. The results of these studies using different values of $\eta$ are given in table ~\ref{tab:table1}. 
\begin{table}
\caption{\label{tab:table1}Obtained values of the parameters  for differents ``data'' samples}
\begin{ruledtabular}  
\begin{tabular}{lccc}
\hline
$\eta$& $\alpha$ & $\beta$ & $\gamma$  \\
\hline
-0.4  & 0.002 $\pm$ 0.03 & -0.05 $\pm$ 0.02 & 0.98  $\pm$ 0.04 \\
-0.25 & 0.08  $\pm$ 0.04 & -0.06 $\pm$ 0.02 & 0.92  $\pm$ 0.04 \\
-0.1  & 0.43  $\pm$ 0.04 & -0.09 $\pm$ 0.02 & 0.57  $\pm$ 0.04 \\
-0.05 & 0.69  $\pm$ 0.04 & -0.06 $\pm$ 0.02 & 0.31  $\pm$ 0.04 \\ 
0     & 0.97  $\pm$ 0.05 &  0.003$\pm$ 0.02 & 0.03  $\pm$ 0.04 \\
0.05  & 0.70  $\pm$ 0.05 &  0.05 $\pm$ 0.02 & 0.29  $\pm$ 0.04 \\
0.1   & 0.40  $\pm$ 0.04 &  0.04 $\pm$ 0.02 & 0.59  $\pm$ 0.04 \\
0.25  & 0.08  $\pm$ 0.04 &  0.04 $\pm$ 0.02 & 0.92  $\pm$ 0.04 \\
0.4   & 0.002 $\pm$ 0.03 &  0.01 $\pm$ 0.02 & 0.98  $\pm$ 0.04 \\
\hline 
\end{tabular}
\end{ruledtabular}
\end{table}

The value of $\alpha$ gives the fraction of the scalar $J^{PC}= 0^{++}$ component of the Higgs boson, while $\gamma$ gives the contribution of the pseudoscalar Higgs component and increases quickly with $\eta$  as expected. It can be seen from our results that the  Monte Carlo study using a sample
of pure scalar SM-like Higgs gives a consistent answer. This indicates the high sensitivity of the method to distinguish a purely $\cal CP$-even state from a pseudoscalar $\cal CP$-odd state. Secondly, the method also allows one to determine whether the observed Higgs boson is a $\cal CP$ mixture and, if so, measure the odd and even component. It is evident that the statistical uncertainties prevent us to a large extent from measuring the interference term. It should be noted that for  $Z\rightarrow \mu^{+}\mu^{-}$, as well as for $Z\rightarrow e^{+}e^{-}$, the interference term is suppressed by the smallness of $v_f$ independently of the size of $\eta$. However, the simultaneous existence of fractions $\alpha$ and $\gamma$ would indicate $\cal CP$ violation for the $ZZ\phi$ coupling. The method proposed here gives sensible results in the case that there is any significant $\cal CP$-even component in the $\phi$ Higgs boson or if  $\phi$ is almost purely $\cal CP$-odd. The statististical significance can certainly 
be increased including the  $e^+ e^- \rightarrow e^+ e^- X$ channel.

\section{Summary}

We have proposed a novel method for the measurement of the parity 
of the Higgs boson using the angular distributions of the 
differential cross section of $e^+e^- \rightarrow Z \phi$. The statistical power of our method using Monte Carlo generated hypothetical data samples is shown in Table ~\ref{tab:table1}. The results indicate that, for an integrated luminosity of $500 fb^{-1}$, at 350 GeV centre-of-mass energy, TESLA will be able to unambiguosly determine whether a Higgs boson is a state $0^{++}$ ($\cal CP$-even, scalar) or has a contribution of the  $0^{-+}$ ($\cal CP$-odd, pseudoscalar) state, like in
general extensions of Higgs model. We also estimate the statistical uncertainties for the measurement of the $\cal CP$ violating interference term. We hope that this technique will allow confirmation of the expected   $J^{PC}$ assignment of a Higgs boson candidate.

\section{Acknowledgments}

We are greateful to W. Lohmann with whom early aspects of this idea were discussed. We would also like to thank L. Epele  and P.Garcia-Abia for helpful discussions and T. Paul for carefully reading the manuscript. This work has been supported, in part by CONICET, Argentina. MTD thanks the John Simon Guggenheim Foundation for a fellowship.

\end{document}